\documentclass[12pt]{iopart}                  
\newcommand{\text}[1]{\hbox{#1}}              
\newcommand{\bss}{\mathbf}                    
\newcommand{\D}{\rmd}                         
\newcommand{\gr}{{\rm\scriptstyle g}}
\newcommand{\g}{\hbox{\sl g}}
\newcommand{\Cdot}{\,}
\newcommand{\A}{\hbox{$\cal A$}}
\newcommand{\E}{\hbox{$\rm E$}}
\newcommand{\R}[1]{{\bi #1}}
\newcommand{\B}[1]{\protect{\bss #1}}
\newcommand{\AAA}{\protect{\bss A}}
\newcommand{\VV}{\protect{\bss V}}
\newcommand{\UU}{\protect{\bss U}}
\newcommand{\AV}{\hbox{$\AAA\Cdot\VV$}}
\newcommand{\VA}{\hbox{$\VV\Cdot\AAA$}}
\newcommand{\kE}{k_{\scriptstyle \rm E}}
\begin{document}

\title{Accelerated motion and special relativity transformations}

\author{A A Ketsaris\footnote{{\it E-mail address\/}:
ketsaris@sai.msu.su}}
\address{19-1-83, ul. Krasniy Kazanetz, Moscow 111395, Russian Federation}
\date{\today}
\jl{1}

\begin{abstract}

Transformation rules for coordinates,
velocities and accelerations in accelerated reference frames
are derived.
A generalized approach of the special relativity is taken for a basis.
A 7-dimensional space including projections of velocity vector as three
additional coordinates to time and geometric coordinates is studied.
Turns in pseudoplane $(\D t,\D v)$ of this 7-space describe accelerated
motion of frame.
In addition to the light velocity, the transformation formulas contain
a fundamental constant which has a meaning of maximal acceleration.
It is demonstrated that if a source of light moves with acceleration
with respect to some reference frame, the light velocity in this frame is
less than the constant $c$ and depends on acceleration.
The special relativity relation between energy, impulse, and mass gets
changed for particle in accelerated motion.
A generalized wave operator being invariant to the above
transformations is introduced.
The components of tensor and of potential of electromagnetic field
get intermixing in transformation relations for accelerated frame.
\end{abstract}

\section{Introduction}

At the present time the Special Relativity Theory (SRT) is widely used for
the {\it approximate} description of motion in "non-fully-inertial" reference
frames.  Yet, there is a number of relativistic phenomena defying analysis in
the framework of the SRT. In particular, the SRT allow to consider neither
the process of formation (emission) of photon as particle moving with light
velocity, nor its proper rotation.  Moreover composition rules for
accelerations, angular velocities, and for other kinematic parameters
except velocities can in no
way be found from the SRT.  This problem has its origin in rough kinematic
notions of light which form the basis of SRT.

The necessity of generalizing the SRT stems also from
mathematical reasons. The matter is that a relativistic approach now-used
for accelerated reference frames has the grave logical disadvantage. The
additional algebraic relation between velocity and a turn angle, $\psi$, in
pseudoplane $(x,t)$
$$
     \frac{\D x}{\D t} = c\Cdot\tanh \psi
$$
is introduced in the SRT.  In particular, this relation allows
to derive the relativistic composition rule for velocities.  Yet, a similar
algebraic relation is not considered for acceleration.  Thus the uniformity
of the description of different kinematic parameters is missing.
For this reason, generalized relativistic relations for acceleration can
escape our understanding and the composition rule for accelerations being
similar to that of velocities can not be formulated.

The objective of this work is to develop a SRT generalization
to accelerated motions from the principle of the uniform description of
acceleration and of velocity. Adhering to this heuristic principle,
one can achieve
the essential progress in describing the kinematic
properties of light. Further it will be shown that
\begin{enumerate}

\item
light can move with acceleration;

\item
there is a maximal acceleration of light, $A$;

\item
the velocity, $v$, and acceleration, $a$, of light follow to the relation
\[
   c^2 - v^2 - \frac{c^2}{A^2} a^2 = 0\,;
\]

\item
the specified relation and the fundamental constants
entered into it, the velocity $c$ and the acceleration $A$, do not depend on
velocity and acceleration of reference frame.

\end{enumerate}
Note that the proposal of the existence of maximal acceleration is not
radically new.  It was first made by Caianiello in 1981 \cite{Cai81} in the
context of a quantization model formulated in an eight-dimensional geometric
phase space, with coordinates $x^\alpha=\{x^i,(\hbar/mc)\Cdot u^i\}$, where
$x^i$ is the position four-vector, and $u^i=\D x^i/\D s$ is the relativistic
four-velocity $(i=1,\ldots,4)$.

\section{Formulation of problem}

Consider a body $B$ in linear motion with respect to a frame $K$
at velocity $v$ and acceleration $a$.  Let a frame $K'$ move linear
with respect to $K$ at velocity $\R{v}$ and acceleration $\R{a}$. Let the
motion of $B$ be characterized by velocity $v'$ and acceleration $a'$
with respect to $K'$.

In the Newtonian mechanics, the kinematics of linear uniform accelerated
motion is described by the system of differential equations
\begin{equation}
\eqalign{
     \D t   &=            \D t'         \,,\\
     \D x   &= \R{v}\Cdot \D t' + \D x' \,,\\
     \D v   &= \R{a}\Cdot \D t' + \D v' \,,\\
     \D a   &= \D a' = 0                \,.}
\label{F1}
\end{equation}

We assume that the kinematics of linear uniform accelerated motion,
in the generalized SRT desired us, is described by a system
\begin{equation}
\eqalign{
   \begin{array}{||c||}
    \D t \\ \D x \\ \D v
   \end{array}
   = \B{F}\Cdot
   \begin{array}{||c||}
    \D t' \\ \D x' \\ \D v'
   \end{array}  \,,\\
   \D a = \D a' = 0 \,,}
\label{F2}
\end{equation}
where the elements of transformation matrix $\B{F}$ are functions of
velocity and of acceleration.  For linear uniform motion, the
specified system of differential equations should be reduced to the Lorentz
transformations. Therefore
\[
     \B{F}(\R{v},0) =
     \left(
     \begin{array}{c|c|c}
          \frac{1}{\sqrt{1-\frac{\R{v}^2}{c^2}}} &
          \frac{\R{v}}{c^2\Cdot\sqrt{1-\frac{\R{v}^2}{c^2}}} &  0 \\
     \hline
          \frac{\R{v}}{\sqrt{1-\frac{\R{v}^2}{c^2}}} &
          \frac{1}{\sqrt{1-\frac{\R{v}^2}{c^2}}}             &  0 \\
     \hline
           0 & 0 & 1 \\
     \end{array}
     \right) \,.
\]
Moreover at the Newtonian limit it should be fulfilled
\[
     \B{F}(\R{v},\R{a}) =
     \left(
     \begin{array}{c|c|c}
          1     & 0 & 0 \\
     \hline
          \R{v} & 1 & 0 \\
     \hline
          \R{a} & 0 & 1 \\
     \end{array}
     \right) \,.
\]
In a special case, the relations (\ref{F2}) should also reduce to the Rindler
transformations (see, e.g., \cite{Rin,Mis})
\begin{equation}
     t = \frac{c}{a}\Cdot \sinh\left(\frac{a\Cdot t'}{c}\right)\,,  \qquad
     x = \frac{c^2}{a}\Cdot \cosh\left(\frac{a\Cdot t'}{c}\right) \,.
\label{F2.5}
\end{equation}

The desired generalization of the SRT must meet to the
specified requirements.

A common approach for the necessary generalizations is developed in
Section~3 by using the standard SRT as an example.  The accelerated
motion kinematics, including composition rules for velocities and
accelerations and transformations of kinematic variables, is considered
in Section~4.  In particular, the reduction to the Rindler
transformations is produced.  Section~4 contains, in addition,
transformations of potential and of tensor of electromagnetic field for
accelerated frames.  A generalization of dynamic variables (energy,
impulse, force) is produced in Section~5 where the standard wave equation
is also modified to describe accelerated motion of wave.  The
conclusions are presented in Section~6.

\section{Lorentz transformations: a generalized approach}

In this Section the Lorentz transformations 
will be derived by a new original method which will be important for further
analysis.

An interval square used in the SRT can be represented as
\[
     (\D s)^2 = c^2\Cdot (\D t)^2 - (\D x)^2  \,.
\]
The variables are changed conveniently by:
\[
     \D x^1 = \D x \,,\qquad
     \D x^4 = c\Cdot \D t \,.
\]
Using the new variables we have the interval square in form
\[
     (\D s)^2 = (\D x^4)^2 - (\D x^1)^2 \,.
\]

The linear transformation preserving the interval square
\begin{equation}
     ||\D x|| = \UU\Cdot ||\D x'||
\label{F8.5}
\end{equation}
is a {\it turn\/} in $(\D x^1,\D x^4)$ pseudoeuclidean plane \cite{Min}.
From the transformation matrix
\[
     \UU \equiv \VV =
     \left(
     \begin{array}{c|c}
     \cosh\Psi & \sinh\Psi \\
     \hline
     \sinh\Psi & \cosh\Psi \\
     \end{array}
     \right)
\]
follows
\begin{equation}
\eqalign{
     \D x^4 &= \cosh\Psi \Cdot (\D x^4)' + \sinh\Psi \Cdot (\D x^1)' \,,\\
     \D x^1 &= \sinh\Psi \Cdot (\D x^4)'+ \cosh\Psi \Cdot (\D x^1)'  \,,}
\label{F3}
\end{equation}
where $\Psi$ is the turn angle.

To find a relation between the angle $\Psi$ and the velocity $x^1_4$ we
consider the variation of coordinate differentials as
function of the angle variation:
\[
     \delta ||\D x|| = \delta \UU\Cdot ||\D x'|| \,.
\]
Taking into account that
\[
     ||\D x'|| = \UU^{-1}\Cdot ||\D x||
\]
we can set
\begin{equation}
     \delta ||\D x|| = (\delta \UU\Cdot \UU^{-1})\Cdot ||\D x|| \,.
\label{F9}
\end{equation}
In our case this relation is reduced to
\begin{equation}
\eqalign{
     \delta \D x^4 &= \delta \Psi \Cdot \D x^1  \,,\\
     \delta \D x^1 &= \delta \Psi \Cdot \D x^4  \,.}
\label{F4}
\end{equation}

Now let us consider the differential
\begin{equation}
     \delta \frac{\D x^1}{\D x^4} = \frac{\delta \D x^1}{\D x^4} -
     \D x^1 \Cdot \frac{\delta \D x^4}{(\D x^4)^2} \,.
\label{F5}
\end{equation}
Using (\ref{F4}) we find
\begin{equation}
     \delta \frac{\D x^1}{\D x^4} = \delta \Psi\Cdot
     \left[ 1 - \left(\frac{\D x^1}{\D x^4} \right)^2 \right] \,.
\label{F6}
\end{equation}
The relations \eref{F3} and \eref{F6} describe the Lorentz transformations and
the SRT composition rule for velocities.

The principal positions of the mathematical approach used above
must be emphasized:
\begin{enumerate}
\item
A vector space is built on the differentials of
function $x(t)$ and of argument $t$.

\item
The linear transformations preserving the interval square
of this space are considered.

\item
A correspondence exists between the said linear transformations and the
derivative $v=\D x/\D t$.

\item
The rule of composition of the linear transformations corresponds to the
composition rule in the space of derivatives (i.e. the velocity composition
rule).

\end{enumerate}
We assume that this approach has an invariant meaning and can be
applied to the function $v(t)$. This point of view allows to obtain
transformations for accelerated frame and the composition rule for
accelerations in according with the special relativity principles.

\section{Transformations for accelerated frames}

In addition to the function $x(t)$ we consider the function
$v(t)=\D x/\D t$. Correspondingly to the previous considerations we
construct a vector space on differentials $\D x$, $\D t$, $\D v$
and set the interval square in the following form
\[
     (\D s)^2 = c^2\Cdot (\D t)^2 - (\D x)^2 -  T^2\Cdot (\D v)^2 \,.
\]
Here the constant $T$ adjusts the dimensionality of $\D v$ to the
dimensionality of interval and has the dimensionality of time. Setting
the constant
\[
     L = c\Cdot T
\]
with the dimensionality of length and dividing the interval square
by $L^2$ we obtain
\[
     (d\sigma)^2 \equiv \frac{(\D s)^2}{L^2} = \frac{(\D t)^2}{T^2} -
     \frac{(\D x)^2}{L^2} - \frac{(\D v)^2}{c^2} \,.
\]
In this expression all components are dimensionless.
Let us change variables
\[
     x^1 = \frac{x}{L}\,,\qquad x^4 = \frac{t}{T}\,,\qquad
     x^1_4 = \frac{v}{c} = \frac{\D x^1}{\D x^4}\,.
\]
Then the interval square is
\begin{equation}
     (d\sigma)^2 = (\D x^4)^2 - (\D x^1)^2 - (\D x^1_4)^2 \,.
\label{F7}
\end{equation}

Further we shall make considerations about a sign of square of velocity
differential in the interval square. One can define the dimensionless
interval square through covariant and contravariant vector coordinates
as
\[
    - (d\sigma)^2 = \D x^4\Cdot \D x_4 + \D x^1\Cdot \D x_1 +
    {{\D x^1}_4}\Cdot {{\D x^4}_1} \,.
\]
Here all addends on the right have a positive sign.
If we use by relations
\[
     \D x_1 = \g_{11} \Cdot \D x^1 = \D x^1  \,,\qquad
     \D x_4 = \g_{44} \Cdot \D x^4 = - \D x^4\,,\qquad
     {{\D x^4}_1} = {{\D x^1}_4} \,,
\]
where it is taken into account that
the turn matrix in pseudoeuclidean plane does not change sign by
transposition, we get (\ref{F7}).
Thus the coordinate $x^1_4$ is {\it spacelike}.

Let us introduce the following notations for dimensionless motion
parameters:  $x^1$, $x^4$, $x^1_4$, $x^1_{44}$ as coordinate, time,
velocity and acceleration of body $B$ with respect to frame $K$;
$(x^1)'$, $(x^4)'$, $(x^1_4)'$, $(x^1_{44})'$ as that with respect to
$K'$; and $\R{x}^1$, $\R{x}^4$, $\R{x}^1_4$, $\R{x}^1_{44}$ as that for
frame $K'$ with respect to $K$.

A turn in the space of differentials
\[
     ||\D x|| = \UU\Cdot ||\D x'||
\]
preserves the interval square.
The turns in pseudoplanes $(\D x^1,\D x^4)$ and
$(\D x^4,\D x^1_4)$:
\[
     \VV =
     \left(
     \begin{array}{c|c|c}
     \cosh\Psi & \sinh\Psi & 0 \\
     \hline
     \sinh\Psi & \cosh\Psi & 0 \\
     \hline
      0        & 0         & 1 \\
     \end{array}
     \right) \,,\qquad
     \AAA =
     \left(
     \begin{array}{c|c|c}
     \cosh\Phi & 0  & \sinh\Phi \\
     \hline
      0        & 1  & 0         \\
     \hline
     \sinh\Phi & 0  & \cosh\Phi \\
     \end{array}
     \right)
\]
describe uniform velocity and accelerated motions of frame, respectively.
Since turns are in general {\it non-commuting}, a motion with
$\UU=\VA$ must be distinguished from a motion with $\UU=\AV$.

\subsection{\VA-motion}

Consider turns described by matrix $\UU=\VA$.
The relations \eref{F8.5} and \eref{F9} take form:
\begin{equation}
\eqalign{
     \D x^4   &= \cosh\Psi\Cdot \cosh\Phi \Cdot (\D x^4)' +
     \sinh\Psi \Cdot (\D x^1)' +
     \cosh\Psi\Cdot \sinh\Phi \Cdot (\D x^1_4)' \,,\\
     \D x^1   &= \sinh\Psi\Cdot \cosh\Phi \Cdot (\D x^4)' +
     \cosh\Psi \Cdot (\D x^1)' +
     \sinh\Psi\Cdot \sinh\Phi \Cdot (\D x^1_4)' \,,\\
     \D x^1_4 &= \sinh\Phi \Cdot (\D x^4)' +
     \cosh\Phi \Cdot (\D x^1_4)'      \,,}
\label{F8}
\end{equation}
and
\begin{equation}
\eqalign{
     \delta \D x^4   &= \delta \Psi \Cdot \D x^1 +
     \cosh\Psi\Cdot \delta \Phi\Cdot \D x^1_4  \,,\\
     \delta \D x^1   &= \delta \Psi \Cdot \D x^4 +
     \sinh\Psi\Cdot \delta \Phi\Cdot \D x^1_4  \,,\\
     \delta \D x^1_4 &= \cosh\Psi\Cdot \delta \Phi\Cdot \D x^4 -
     \sinh\Psi\Cdot \delta \Phi\Cdot \D x^1    \,.}
\label{F10}
\end{equation}

To find a relation connecting the angles $\Psi$ and $\Phi$
with the velocity $x^1_4$ we consider the
differential $\delta x^1_4$.
Using (\ref{F5}) and (\ref{F10}) we obtain
\begin{equation}
     \delta x^1_4 = \left[ 1 - (x^1_4)^2 \right] \Cdot \delta\Psi
     +
     (\sinh\Psi - x^1_4\Cdot \cosh\Psi)\Cdot x^1_{44}
     \Cdot \delta\Phi \,.
\label{F11}
\end{equation}

To find a relation connecting the angles $\Psi$ and $\Phi$ with the
acceleration $x^1_{44}$ we consider the differential
\[
     \delta \frac{\D x^1_4}{\D x^4} = \frac{\delta \D x^1_4}{\D x^4} -
     \D x^1_4 \Cdot \frac{\delta \D x^4}{(\D x^4)^2} \,.
\]
Using (\ref{F10}) we obtain
\begin{equation}
     \delta x^1_{44} = - x^1_{44}\Cdot x^1_4\Cdot \delta \Psi
     +
     \left\{\cosh\Psi\Cdot \left[ 1 -(x^1_{44})^2 \right] -
     \sinh\Psi\Cdot x^1_4 \right\}\Cdot \delta \Phi\,.
\label{F12}
\end{equation}

The relations (\ref{F8}), (\ref{F11}) and (\ref{F12}) describe
the coordinate transformations for the accelerated motion
of body $B$ and of frame $K'$ as well as the composition rules for
velocities and accelerations.

Further we shall find a solution of equations (\ref{F11}) and
(\ref{F12}), when the velocity of body $B$ is zero with respect to $K'$.
In this case $v=\R{v}$.

\subsubsection{Composition rule for accelerations}

From (\ref{F11}) under initial condition ($\Psi = 0$, $x^1_4 = 0$)
follows
\begin{equation}
     x^1_4 = \R{x}^1_4 = \tanh \Psi\,.
\label{F13}
\end{equation}
If the velocity $x^1_4=0$ and $\Psi=0$ then
(\ref{F12}) gets reduced to
\[
     \delta x^1_{44} =
     \left[1 - (x^1_{44})^2 \right]\Cdot \delta \Phi\,.
\]
The solution of this equation is
\begin{equation}
     x^1_{44} = \tanh(\Phi+\phi')=\tanh\phi \,,
\label{F14}
\end{equation}
where $\phi'$ is integration constant and notation $\phi=\Phi+\phi'$
is used. The integration constant $\phi'$ can be found from the following
considerations.  Let us consider that $\phi'=0$ corresponds $a'=0$, and
$\Phi=0$ corresponds $\R{a}=0$. Thus
\[
     \R{x}^1_{44} = \tanh\Phi\,, \qquad (x^1_{44})' = \tanh\phi' \,.
\]
Hereof and from (\ref{F14}) the composition rule for accelerations follows
\[
     x^1_{44} =
     \frac{\R{x}^1_{44} + (x^1_{44})'}{1+\R{x}^1_{44}\Cdot (x^1_{44})'} \,.
\]
Changing for dimensional acceleration in accordance with
\[
     x^1_{44} = \frac{\D x^1_4}{\D x^4} =
     \frac{T}{c}\Cdot \frac{\D v}{\D t} = \frac{a}{A} \,,
\]
where the constant with dimensionality of acceleration
\[
     A=\frac{c}{T}
\]
was introduced, we obtain
\[
     a = \frac{\R{a}+ a'}{1+\frac{\R{a}\Cdot a'}{A^2}} \,.
\]
Thus the resulting acceleration is always less than or equal to $A$.
Note that the above composition rule for accelerations was derived by
Scarpetta within the framework of Caianiello's model \cite{Sca}.

At non-zero velocity $x^1_4$ with using (\ref{F13})
the equation (\ref{F12}) yields
\begin{equation}
     x^1_{44} = \sqrt{1- (\R{x}^1_4)^2} \Cdot \tanh(\Phi+\phi')\,.
\label{F15}
\end{equation}
Because of correspondence between $\phi'$ and $a'$, $\Phi$ and
$\R{a}$ we have
\begin{equation}
     \R{x}^1_{44} = \sqrt{1-(\R{x}^1_4)^2}\Cdot \tanh \Phi
\label{F16}
\end{equation}
and $(x^1_{44})' = \tanh \phi'.$
In the last relation it is taken into account that $(x^1_4)'=0$.
Hereof and from (\ref{F15}) the general composition rule for accelerations
follows
\[
     a = \sqrt{1-\frac{\R{v}^2}{c^2}}\Cdot
     \frac{\R{a} + \sqrt{1-\frac{\R{v}^2}{c^2}}\Cdot a'}{
     \displaystyle
     \sqrt{1-\frac{\R{v}^2}{c^2}}+\frac{\R{a}\Cdot a'}{A^2}} \,.
\]

\subsubsection{Transformation of coordinate differentials}

From (\ref{F13}) and (\ref{F16}) follows
\[
     \cosh\Psi = \frac{1}{\sqrt{1-(\R{x}^1_4)^2}} \,, \qquad
     \sinh\Psi = \frac{\R{x}^1_4}{\sqrt{1-(\R{x}^1_4)^2}} \,,
\]
\[
     \cosh\Phi = \case{\sqrt{1-(\R{x}^1_4)^2}}%
     {\sqrt{1-(\R{x}^1_4)^2-(\R{x}^1_{44})^2}} \,, \qquad
     \sinh\Phi =
     \case{\R{x}^1_{44}}{\sqrt{1-(\R{x}^1_4)^2-(\R{x}^1_{44})^2}} \,.
\]
Substituting the above expressions in (\ref{F8}) we obtain
the transformations of coordinate differentials:
\begin{equation}
\fl
\eqalign{
     \D x^4   &= \case{1}{\sqrt{1-(\R{x}^1_4)^2-(\R{x}^1_{44})^2}}
     \Cdot (\D x^4)'
     {}+ \case{\R{x}^1_4}{\sqrt{1-(\R{x}^1_4)^2}} \Cdot (\D x^1)'
     {}+
     \case{1}{\sqrt{1-(\R{x}^1_4)^2}}\Cdot
     \case{\R{x}^1_{44}}{\sqrt{1-(\R{x}^1_4)^2-(\R{x}^1_{44})^2}}
     \Cdot (\D x^1_4)'
     \,,\\
     \D x^1   &= \case{\R{x}^1_4}{\sqrt{1-(\R{x}^1_4)^2-(\R{x}^1_{44})^2}}
     \Cdot (\D x^4)'
     {}+ \case{1}{\sqrt{1-(\R{x}^1_4)^2}} \Cdot (\D x^1)'
     {}+
     \case{\R{x}^1_4}{\sqrt{1-(\R{x}^1_4)^2}}\Cdot
     \case{\R{x}^1_{44}}{\sqrt{1-(\R{x}^1_4)^2-(\R{x}^1_{44})^2}}
     \Cdot (\D x^1_4)'
     \,,\\
     \D x^1_4 &= \case{\R{x}^1_{44}}{\sqrt{1-(\R{x}^1_4)^2-(\R{x}^1_{44})^2}}
     \Cdot (\D x^4)'
     {}+
     \case{\sqrt{1-(\R{x}^1_4)^2}}{\sqrt{1-(\R{x}^1_4)^2-(\R{x}^1_{44})^2}}
     \Cdot (\D x^1_4)' \,.}
\label{F17}
\end{equation}
Thus we have found a concrete form of the transformations (\ref{F2}).
By using dimensional values for $\R{v}\ll c$ and $\R{a}\ll A$ the
transformations are reduced to
\begin{eqnarray*}
     \D t   &=& \D t' +\case{1}{c^2}\Cdot \R{v}\Cdot \D x' +
              \case{1}{A^2}\Cdot \R{a}\Cdot \D v'        \,,\\
     \D x   &=& \R{v}\Cdot \D t' + \D x' +
              \case{1}{A^2}\Cdot \R{a}\Cdot \R{v}\Cdot \D v' \,,\\
     \D v   &=& \R{a}\Cdot \D t' + \D v'                   \,.
\end{eqnarray*}
At the Newtonian limit ($c \to \infty$ and $A \to \infty$) we obtain the
system of the differential equations (\ref{F1}).

\subsection{\AV-motion}

We now consider the case when turns are described by matrix $\UU =\AV$.
The coordinate transformations and the composition rules for velocities
and accelerations can be obtained, much as it was made in the previous
Section.  We give results for $a'=0$.

The velocity composition rule has form
\[
     v = \sqrt{1-\frac{\R{a}^2}{A^2}}\Cdot
     \frac{\R{v} + \sqrt{1-\frac{\R{a}^2}{A^2}}\Cdot v'}{
     \displaystyle
     \sqrt{1-\frac{\R{a}^2}{A^2}}+\frac{\R{v}\Cdot v'}{c^2}} \,.
\]
If the body $B$ is light source, the light speed
with respect to $K$ is calculated by
\[
     v = c\Cdot \sqrt{1-\frac{\R{a}^2}{A^2}} \,.
\]
Therefore the maximal velocity of accelerated motion is {\it less} than $c$.

The transformations of coordinate differentials have form
\begin{equation}
\fl
\eqalign{
     \D x^4   &=
     \case{1}{\sqrt{1-(\R{x}^1_4)^2-(\R{x}^1_{44})^2}} \Cdot (\D x^4)'
     {}+ \case{\R{x}^1_4}{\sqrt{1-(\R{x}^1_4)^2-(\R{x}^1_{44})^2}} \Cdot
     \case{1}{\sqrt{1-(\R{x}^1_{44})^2}} \Cdot (\D x^1)' +
     \case{\R{x}^1_{44}}{\sqrt{1-(\R{x}^1_{44})^2}} \Cdot (\D x^1_4)'
     \,,\\
     \D x^1   &=
     \case{\R{x}^1_4}{\sqrt{1-(\R{x}^1_4)^2-(\R{x}^1_{44})^2}}\Cdot (\D x^4)'
     +
     \case{\sqrt{1-(\R{x}^1_{44})^2}}{\sqrt{1-(\R{x}^1_4)^2-(\R{x}^1_{44})^2}}
     \Cdot (\D x^1)'
     \,,\\
     \D x^1_4 &=
     \case{\R{x}^1_{44}}{\sqrt{1-(\R{x}^1_4)^2-(\R{x}^1_{44})^2}}
     \Cdot (\D x^4)'
     {}+ \case{\R{x}^1_4}{\sqrt{1-(\R{x}^1_4)^2-(\R{x}^1_{44})^2}} \Cdot
     \case{\R{x}^1_{44}}{\sqrt{1-(\R{x}^1_{44})^2}} \Cdot (\D x^1)' +
     \case{1}{\sqrt{1-(\R{x}^1_{44})^2}} \Cdot (\D x^1_4)' \,.}
\label{F18}
\end{equation}
By using dimensional values for $\R{v}\ll c$ and $\R{a}\ll A$ these
transformations are reduced to
\begin{eqnarray*}
     \D t   &=& \D t'+ \case{1}{c^2}\Cdot \R{v}\Cdot \D x' +
              \case{1}{A^2}\Cdot \R{a}\Cdot \D v'                \,,\\
     \D x   &=& \R{v}\Cdot \D t' + \D x'                       \,,\\
     \D v   &=& \R{a}\Cdot \D t' +
              \case{1}{c^2}\Cdot \R{v}\Cdot \R{a}\Cdot \D x' + \D v' \,.
\end{eqnarray*}
Hereof the possibility of passing to the Newtonian kinematics follows also.

\subsection{The special case: Rindler transformations}

Let us derive the Rindler transformations, for example,
from $\VA$-motion transformations. Consider the relations
(\ref{F8}) when
\begin{equation}
     (\D x^1)'= 0 \,, \quad (\D x^1_4)'= 0 \,.
\label{F19}
\end{equation}
In this case $x^1_4 = \R{x}^1_4$, $x^1_{44} = \R{x}^1_{44}$ and
\begin{equation}
\eqalign{
     \D x^4   &= \cosh\Psi\Cdot \cosh\Phi \Cdot (\D x^4)' \,,\\
     \D x^1   &= \sinh\Psi\Cdot \cosh\Phi \Cdot (\D x^4)' \,,\\
     \D x^1_4 &= \sinh\Phi \Cdot (\D x^4)'\,.}
\label{F20}
\end{equation}
Here the angles $\Psi$ and $\Phi$ are related with velocity and
acceleration of frame $K'$ by (\ref{F13}) and (\ref{F16}).
Let these angles be small and
\[
     \Psi \approx x^1_4  \,, \qquad
     \sinh\Phi \approx \tanh \Phi \approx x^1_{44} \,, \qquad
     \cosh\Phi \approx 1\,.
\]
Then the transformations (\ref{F20}) are reduced to
\begin{eqnarray*}
     \D x^4   &=& \cosh(x^1_4)\Cdot (\D x^4)' \,,\\
     \D x^1   &=& \sinh(x^1_4)\Cdot (\D x^4)' \,,\\
     \D x^1_4 &=& x^1_{44} \Cdot (\D x^4)'\,.
\end{eqnarray*}
If we integrate these equations under the constancy of
acceleration $x^1_{44}$ and assume integration constants are zeros, we
obtain
\[
     x^4 = \frac{1}{x^1_{44}}\Cdot \sinh[x^1_{44} \Cdot (x^4)'] \,,\qquad
     x^1 = \frac{1}{x^1_{44}}\Cdot \cosh[x^1_{44} \Cdot (x^4)']\,.
\]
Hereof the Rindler transformations \eref{F2.5} follow.  Note that the
time $t'$ is proper time $\tau=s/c$ in the Rindler transformations.
Really, from (\ref{F19}) follows
\[
      (\D s)^2 = c^2\Cdot (\D t')^2\,.
\]

\subsection{Transformations for potential and tensor of electromagnetic field}

Transformation formalism for potential and tensor of electromagnetic field
follows from (\ref{F17}) and (\ref{F18}) with due regard that potential,
$\A^i$, gets transformed similar to $\D x^i$, tensor components, $F^b_4$, get
transformed similar to velocity $\D x^b_4$, and sets of components
$\left\{F^4_2\,,\;F^1_2\right\} =\left\{-\E^2\,,\;B^3\right\}$,
$\left\{F^4_3\,,\;F^1_3\right\} =\left\{-\E^3\,,\;-B^2\right\}$
get transformed similar to $\{\D x^4\,,\; \D x^1\}$ components.
Because the transformations (\ref{F17}) and (\ref{F18}) do not
affect on components $\D x^2$ and $\D x^3$ then components
$\A^2$, $\A^3$ and $F^2_3=B^1$ remain constant.

In particular for \AV-motion, the transformations have form
\[
\fl
\eqalign{
     \varphi   &=
     \case{1}{\sqrt{1-(\R{x}^1_4)^2-(\R{x}^1_{44})^2}} \Cdot (\varphi)'
     {}+  \case{\R{x}^1_4}{\sqrt{1-(\R{x}^1_4)^2-(\R{x}^1_{44})^2}} \Cdot
     \case{1}{\sqrt{1-(\R{x}^1_{44})^2}} \Cdot (\A^1)'
     + \case{\R{x}^1_{44}}{\sqrt{1-(\R{x}^1_{44})^2}} \Cdot (\E^1)' \,,\\
     \A^1      &=
     \case{\R{x}^1_4}{\sqrt{1-(\R{x}^1_4)^2-(\R{x}^1_{44})^2}}
     \Cdot (\varphi)' +
     \case{\sqrt{1-(\R{x}^1_{44})^2}}{\sqrt{1-(\R{x}^1_4)^2-(\R{x}^1_{44})^2}}
     \Cdot (\A^1)' \,,\\
     \E^1       &=
     \case{\R{x}^1_{44}}{\sqrt{1-(\R{x}^1_4)^2-(\R{x}^1_{44})^2}}
     \Cdot (\varphi)'
     {}+ \case{\R{x}^1_4}{\sqrt{1-(\R{x}^1_4)^2-(\R{x}^1_{44})^2}} \Cdot
     \case{\R{x}^1_{44}}{\sqrt{1-(\R{x}^1_{44})^2}} \Cdot (\A^1)' +
     \case{1}{\sqrt{1-(\R{x}^1_{44})^2}} \Cdot (\E^1)' \,,\\
     B^3       &=
     - \case{\R{x}^1_4}{\sqrt{1-(\R{x}^1_4)^2-(\R{x}^1_{44})^2}}\Cdot (\E^2)'
     + \case{\sqrt{1-(\R{x}^1_{44})^2}}{\sqrt{1-(\R{x}^1_4)^2
          -(\R{x}^1_{44})^2}}
     \Cdot (B^3)' \,,\\
     \E^2       &=
     \case{1}{\sqrt{1-(\R{x}^1_4)^2-(\R{x}^1_{44})^2}} \Cdot (\E^2)'
     {}- \case{\R{x}^1_4}{\sqrt{1-(\R{x}^1_4)^2-(\R{x}^1_{44})^2}} \Cdot
     \case{1}{\sqrt{1-(\R{x}^1_{44})^2}} \Cdot (B^3)' \,,\\
     B^2       &=
     \case{\R{x}^1_4}{\sqrt{1-(\R{x}^1_4)^2-(\R{x}^1_{44})^2}}\Cdot (\E^3)' +
     \case{\sqrt{1-(\R{x}^1_{44})^2}}{\sqrt{1-(\R{x}^1_4)^2-(\R{x}^1_{44})^2}}
     \Cdot (B^2)' \,,\\
     \E^3       &=
     \case{1}{\sqrt{1-(\R{x}^1_4)^2-(\R{x}^1_{44})^2}} \Cdot (\E^3)'
     {}+ \case{\R{x}^1_4}{\sqrt{1-(\R{x}^1_4)^2-(\R{x}^1_{44})^2}} \Cdot
     \case{1}{\sqrt{1-(\R{x}^1_{44})^2}} \Cdot (B^2)' \,.}
\]
Note that in these relations addends proportional to the derivatives of
electromagnetic field tensor are omitted, and all variables are
dimensionless.  The intermixing of potential components with tensor
components distinguishes essentially these transformations from the standard
SRT transformations.

At zero velocity, we obtain
\begin{eqnarray*}
     \fl
     \varphi   &=& \case{1}{\sqrt{1-(\R{x}^1_{44})^2}} \Cdot (\varphi)'+
     \case{\R{x}^1_{44}}{\sqrt{1-(\R{x}^1_{44})^2}} \Cdot (\E^1)' \,,\qquad
     \A^1 = (\A^1)' \,, \\
     \fl
     \E^1       &=&
     \case{\R{x}^1_{44}}{\sqrt{1-(\R{x}^1_{44})^2}}\Cdot (\varphi)'+
     \case{1}{\sqrt{1-(\R{x}^1_{44})^2}} \Cdot (\E^1)' \,,\qquad
     B^2 = (B^2)'  \,, \qquad
     B^3 = (B^3)'  \,,\\
     \fl
     \E^2       &=&
     \case{1}{\sqrt{1-(\R{x}^1_{44})^2}} \Cdot (\E^2)' \,,\qquad
     \E^3    {}= \case{1}{\sqrt{1-(\R{x}^1_{44})^2}} \Cdot (\E^3)'\,.
\end{eqnarray*}
%
Let a point charge, $e$, be in accelerated motion with respect to an observer.
In a proper frame of charge,
\[
     (\varphi)' = \frac{k_{\varphi}\Cdot e}{r} \,,
     \qquad (\A^1)' = 0 \,, \qquad (\E)' = \frac{\kE \Cdot e}{r^2} \,.
\]
The coefficients $k_\varphi$ and $\kE$
are introduced here in order that the scalar potential and the
electric intensity can be dimensionless.
Therefore in the frame of observer at the direction of acceleration,
electric intensity is determined by the following expression:
\[
     \E^1 =
     \frac{x^1_{44}}{\sqrt{1-(x^1_{44})^2}} \Cdot
     \frac{k_{\varphi}\Cdot e}{r} +
     \frac{1}{\sqrt{1-(x^1_{44})^2}} \Cdot \frac{\kE \Cdot e}{r^2} \,.
\]
For $a\ll A$, electric intensity contains two addends, the first one is
proportional to acceleration and varies inversely with distance to charge,
while the second one does not depend from acceleration and is in inverse
ratio with distance square. It is well known result which can be obtained by
means of Lienart-Wiechert potentials (see, e.g., \cite{Lan}). Thus the
transformations derived allow, in particular, to determine field of
accelerated charge without using Maxwell equations.

\section{Relativistic mechanics}

Relativistic mechanics generalized to accelerated motion can be constructed
by analogy with relativistic mechanics being invariant with respect to
uniform velocity motion.

\subsection{7-dimensional velocity}

Differentials included in the expression for the interval square
\[
     \fl
     (\D s)^2 = c^2\Cdot (\D t)^2 - (\D x^1)^2 - (\D x^2)^2 - (\D x^3)^2
     - T^2\Cdot (\D v^1)^2 - T^2\Cdot (\D v^2)^2 - T^2\Cdot (\D v^3)^2
\]
can be considered as coordinates of vector in 7-space.
We have
\[
     \D x^\alpha = \{c\Cdot \D t \,,\; \D x^1 \,,\; \D x^2 \,,\; \D x^3 \,,\;
     T\Cdot \D v^1 \,,\; T\Cdot \D v^2 \,,\; T\Cdot \D v^3\}
\]
for contravariant coordinates of vector and
\[
     \D x_\alpha =
     \{c\Cdot \D t \,,\; - \D x_1 \,,\; - \D x_2 \,,\; - \D x_3 \,,\;\\
     - T\Cdot \D v_1 \,,\; - T\Cdot \D v_2 \,,\; - T\Cdot \D v_3\}
\]
for covariant coordinates of vector.
Using the introduced coordinates we rewrite the interval square in
form
\[
     (\D s)^2 = \D x^\alpha\Cdot \D x_\alpha \,,
     \qquad\qquad(\alpha=1,\ldots,7) \,.
\]
Let us introduce a generalized Lorentz factor
\[
    \gamma =  \left( 1-\frac{v^2}{c^2}-\frac{a^2}{A^2} \right)^{-1/2} \,.
\]
Expressing the interval as
\[
     \D s = \frac{c\Cdot \D t}{\gamma}
\]
we define {\it 7-dimensional velocity} as
\[
     u^\alpha \equiv \frac{\D x^\mu}{\D s} =
     \frac{\partial s}{\partial x_\mu} =
     \left\{ \gamma \,,\; \gamma\Cdot \frac{v^b}{c} \,,\;
             \gamma\Cdot \frac{a^b}{A}
     \right\}
\]
in contravariant coordinates and
\[
     u_\alpha \equiv \frac{\D x_\mu}{\D s} =
     \frac{\partial s}{\partial x^\mu} =
     \left\{ \gamma \,,\; -\gamma\Cdot \frac{v_b}{c} \,,\;
             -\gamma\Cdot \frac{a_b}{A}
     \right\}
\]
in covariant coordinates.
Here $b=1,2,3$, $v^b=v_b$, $a^b=a_b$, $v^b\Cdot v_b=v^2$, $a^b\Cdot a_b=a^2$.
It is obvious that
\begin{equation}
     u^\alpha\Cdot u_\alpha = 1 \,.
\label{F21}
\end{equation}

\subsection{Free particle action}

We define a {\it free particle action} as
\[
     S = - m\Cdot c \int\limits_{s_1}^{s_2} \D s\,,
\]
where the integration is over a line in 7-space, $s_1$, $s_2$ are points of
the specified line, and $m$ is the particle mass.
The action can be expressed as an integral over time
\[
     S = - m\Cdot c^2 \int\limits_{t_1}^{t_2}
     \sqrt{1-\frac{v^2}{c^2} - \frac{a^2}{A^2}} \Cdot \D t\,.
\]

\subsection{Energy, impulse, force}

We define a {\it 7-dimensional impulse} as
\[
     p^\alpha = - \frac{\partial S}{\partial x_\alpha} =
     m\Cdot c \Cdot u^\alpha \qquad \text{and} \qquad
     p_\alpha = - \frac{\partial S}{\partial x^\alpha} =
     m\Cdot c \Cdot u_\alpha
\]
in contravariant and covariant coordinates, respectively.
Let us introduce
a {\it relativistic energy}
\[
     E = \gamma\Cdot m\Cdot c^2 \,,
\]
a {\it relativistic impulse}
\[
     p=\gamma\Cdot m\Cdot v \,,
\]
and a {\it relativistic kinetic force}
\[
     f = \gamma\Cdot m\Cdot a \,.
\]
Using the introduced quantities we can express the components of 7-impulse in
form
\[
     p^\alpha = \left\{\frac{E}{c} \,,\; p^b \,,\; T\Cdot f^b \right\}
\qquad \text{and} \qquad
     p_\alpha = \left\{\frac{E}{c} \,,\; - p_b \,,\; - T\Cdot f_b \right\}
     \,.
\]
From (\ref{F21}) follows
\[
     p^\alpha\Cdot p_\alpha = m^2\Cdot c^2\,.
\]
This can be written as the relation for energy, impulse, kinetic
force, and mass in relativistic mechanics generalized to accelerated motions:
\begin{equation}
     \frac{E^2}{c^2} - p^2 - T^2\Cdot f^2 = m^2\Cdot c^2\,.
\label{F22}
\end{equation}
From it follows that massless accelerated particle is like to
particle in uniform velocity motion with ``effective'' mass
\[
   \frac{1}{c} \Cdot \sqrt{\frac{E^2}{c^2} - p^2} =
   \frac{f}{A} \,.
\]

The transformation of 7-impulse components can be described by the formalism
similar to that for the transformation of coordinate differentials. For
example, in the case of \AV-motion of frame, the transformations (\ref{F18})
imply
\begin{eqnarray*}
\fl
     E   &=&
     \frac{1}{\sqrt{1-\case{\R{v}^2}{c^2}-\case{\R{a}^2}{A^2}}} \Cdot(E)'
     {}+
     \frac{\R{v}}{\sqrt{1-\case{\R{v}^2}{c^2}-\case{\R{a}^2}{A^2}}} \Cdot
     \frac{1}{\sqrt{1-\case{\R{a}^2}{A^2}}} \Cdot (p^1)' +
     \frac{\R{a}\Cdot T^2}{\sqrt{1-\case{\R{a}^2}{A^2}}}\Cdot(f^1)'\,, \\
\bs
\fl
     p^1 &=&
     \frac{\R{v}}{c^2\Cdot\sqrt{1-\case{\R{v}^2}{c^2}-\case{\R{a}^2}{A^2}}}
     \Cdot(E)'+
     \frac{\sqrt{1-\case{\R{a}^2}{A^2}}}{\sqrt{1-\case{\R{v}^2}{c^2}-
          \case{\R{a}^2}{A^2}}}
     \Cdot (p^1)'\,,                                              \\
\fl
     f^1 &=&
     \frac{\R{a}}{c^2\sqrt{1-\case{\R{v}^2}{c^2}-\case{\R{a}^2}{A^2}}}
     \Cdot (E)'
     {}+
     \frac{\R{v}\Cdot \R{a}}{c^2\Cdot
     \sqrt{1-\case{\R{v}^2}{c^2}-\case{\R{a}^2}{A^2}}
     \Cdot\sqrt{1-\case{\R{a}^2}{A^2}}} \Cdot (p^1)' +
     \frac{1}{\sqrt{1-\case{\R{a}^2}{A^2}}}\Cdot (f^1)' \,.
\end{eqnarray*}
Unlike the usual SRT transformations of energy and impulse, the inclusion
of accelerated motion leads to intermixing energy, components of impulse and
of force.

\subsection{Wave equation}

7-dimensional derivative operators are given by
\[
     \fl
     \frac{\partial}{\partial x_\alpha} = \left\{
     \frac{1}{c}\frac{\partial }{\partial t} \,,\;
     -\frac{\partial }{\partial x_b} \,,\;
     -\frac{1}{T}\frac{\partial }{\partial v_b}\right\}
     \qquad \text{and} \qquad
     \frac{\partial }{\partial x^\alpha} = \left\{
     \frac{1}{c}\frac{\partial }{\partial t} \,,\;
     \frac{\partial }{\partial x^b} \,,\;
     \frac{1}{T}\frac{\partial }{\partial v^b} \right\} \,.
\]
Using these operators we get wave equation
\[
     \frac{\partial^2 }{\partial x^\alpha \Cdot \partial x_\alpha} \equiv
     \frac{1}{c^2}\frac{\partial^2 }{\partial t^2}
     -\frac{\partial^2 }{\partial x^2} -
     \frac{1}{T^2}\frac{\partial^2 }{\partial v^2} =0 \,.
\]
This wave equation is different from the traditional one by the last addend
allowing to describe accelerated motion of wave
as well as the processes of wave initiation and disappearance.

Let $\phi(t,x,v)$ be an arbitrary function describing the wave field.
We shall try for a solution of wave equation in form
\begin{equation}
     \phi (t, x, v) = \phi_0 \Cdot\exp
     \left[ i\left(
     \kappa_b \Cdot x^b + \xi_b\Cdot v^b - \omega\Cdot t
     \right) \right] \,,
\label{F23}
\end{equation}
where $\omega$ is the wave frequency, $\kappa_b$ are the coordinates of wave
vector. By analogy $\xi_b$ will be named coordinates of a {\it wave vector of
velocity}.

Substitution of the expression (\ref{F23}) in the wave equation
gives
\begin{equation}
     \frac{\omega^2}{c^2} - \kappa^2 -\frac{\xi^2}{T^2} =0 \,.
\label{F24}
\end{equation}
If we multiply this equation by the Planck constant square and
compare the result to (\ref{F22}) within the framework
of wave-corpuscle duality, we obtain a set of relations:
\[
     E = \hbar\Cdot\omega \,, \qquad
     p = \hbar\Cdot\kappa \,, \qquad
     f = \frac{\hbar}{T^2} \Cdot \xi\,.
\]
The first two are de Broglie relations, while the last one connects
the wave vector of velocity with the relativistic kinetic force.

By analogy with the traditional definition of wave velocity
\[
     v^b \equiv \frac{\partial \omega}{\partial \kappa_b}
     = \frac{\kappa^b}{\omega} \Cdot c^2\,.
\]
we define a {\it wave acceleration\/}
\[
     a^b \equiv \frac{\partial \omega}{\partial \xi_b}
     = \frac{\xi^b}{\omega} \Cdot A^2\,.
\]
Substituting these expressions in (\ref{F24}) we find a relation
\begin{equation}
     c^2 - v^2 - a^2\Cdot T^2 = 0 \,,
\label{F26}
\end{equation}
which will be named an {\it equation of wave motion\/}.
If wave propagates in one direction, integration of the equation of
motion (\ref{F26}) reduces to two solutions for wave velocity:
\begin{eqnarray*}
     v_{I}  &=& \pm c \,,\\
     v_{II} &=& \pm c\Cdot \cos (t/T + C_1) \,,
\end{eqnarray*}
where $C_1$ is integration constant.  Thus two types of wave motion are
possible. In the first case, wave is in linear motion at constant
velocity $c$, and in the second case, wave oscillates along line at
amplitude $L$. At time moments $t/T+C_1=\pi\Cdot n$, where $n$ is
integer, wave velocity is equal $\pm c$ for oscillatory motion, and the
change of motion type is made possible. Then the oscillatory motion of wave
can be transformed into the linear uniform velocity motion and vice versa.
For two-dimensional case, the uniform circular motion is a particular
solution of the equation (\ref{F26}). It is described relations:

for circle radius
\[
     r = \frac{L}{\sqrt{1- \frac{v^2}{c^2}}}\Cdot \frac{v^2}{c^2}\,,
\]

for centripetal acceleration
\[
     a = A\Cdot\sqrt{1- \frac{v^2}{c^2}}\,.
\]

The wave equation generalizing the Klein-Gordon equation for massive
particles to accelerated motion can be written in the following form
\[
     \frac{\partial^2\psi}{\partial x^\alpha \Cdot \partial x_\alpha} +
     \frac{m^2 \Cdot c^2}{\hbar^2} \psi = 0\,.
\]
Substitution of (\ref{F23}) in the wave equation gives
\begin{equation}
     \frac{\omega^2}{c^2} - \kappa^2 -\frac{\xi^2}{T^2} =
     \frac{m^2 \Cdot c^2}{\hbar^2} \,.
\label{F27}
\end{equation}
Let's assume
\[
     \omega = \frac{m\Cdot c^2}{\hbar} + \omega' \,,
\]
where $\omega'\ll {m\Cdot c^2}/{\hbar}$.
For non-relativistic frequency $\omega'$ the equation (\ref{F27}) is reduced
to following
\begin{equation}
    \frac{2\Cdot m\Cdot \omega'}{\hbar} - \kappa^2 -\frac{\xi^2}{T^2} = 0\,.
\label{F28}
\end{equation}
By analogy to the traditional definition of wave packet
{\it group velocity}
\[
     v_{\gr}^b \equiv \frac{\partial \omega'}{\partial \kappa_b} =
     \frac{\hbar}{m} \Cdot \kappa^b
\]
we define a wave packet {\it group acceleration\/}
\[
     a_{\gr}^b \equiv \frac{\partial \omega'}{\partial \xi_b}
     = \frac{\hbar}{m} \Cdot \frac{\xi^b}{T^2} \,.
\]
By using group velocity and acceleration,
(\ref{F28}) is reduced to the expression
\[
     \hbar\Cdot\omega'  = \frac{m\Cdot v_{\gr}^2}{2} +
     \frac{m\Cdot T^2\Cdot a_{\gr}^2}{2}\,,
\]
which is a generalization of known de Broglie's relation.

\section{Conclusions}

In the present paper we have developed the self-consistent SRT generalization
to accelerated motions which is founded only on the principle of the uniform
description of kinematic parameters. This principle is, as
applied to SRT, in the existence of relations between derivations and turn
angles in the space of kinematic variables.

In spite of formal character of initial positions, the generalized SRT
gives far-reaching physical consequences.  In particular, the new
fundamental constant $T$ of time dimensionality should be introduced. This
constant breeds a set of secondary constants: the fundamental length
$L=c\Cdot T$, and the maximal acceleration $A=c/T$.  Unfortunately, the
experimental verification of the relativistic composition rule for
accelerations seems to be as yet impossible because of the extraordinarily
high value of the maximal acceleration (for example, Scarpetta estimated it
at $5\times 10^{53} \text{cm}/\text{s}^2$).

We study accelerated motions in the context differed
radically from Caianiello's conception (see \cite{Rivista} and references
therein).  However our results are similar to those obtained in works of
Caianiello and his colleagues \cite{Sca,Cai90}. By methodical
reasons, we expand space-time through three
coordinates of velocity instead of coordinates of four-velocity used in
Caianiello's model.  This explains the difference between coordinate
transformations, and between relations for kinematic and dynamic parameters
in our and Scarpetta's considerations.


Since in according to the SRT, three velocity components and three space turn
angles make full Lorentz group, the present work can be naturally
generalized to uniform rotations of reference frames.  An universal
"relativistic" approach to any non-inertial motions and, in particular, to
uniform rotations will be given in a forthcoming paper \cite{Ket}.



\Bibliography{10}

\bibitem{Rin} Rindler W 1991 {\it Introduction to Special Relativity}
(Oxford: Clarendon) p~33.

\bibitem{Mis} Misner C, Torn K, Wheeler J 1973 {\it Gravitation}
(San Francisco: Freeman and Company) ch~6.

\bibitem{Cai81} Caianiello E R 1981 {\it Lett. Nuovo Cimento} {\bf 32} 65.


\bibitem{Min} Minkowski H 1908 {\it Goett. Nachr.} p~53.

\bibitem{Sca} Scarpetta G 1984 {\it Lett. Nuovo Cimento} {\bf 41} 51.

\bibitem{Lan} Landau L D, Lifshits E M 1967 {\it Field theory\/}
(Moscow: Nauka) p~212.

\bibitem{Rivista} Caianiello E R 1992 {\it La rivista del Nuovo Cimento}
{\bf 15} no~4.

\bibitem{Cai90} Caianiello E R, Feoli A, Gasperini M, Scarpetta G 1990
{\it Int.~J.~Theor.~Phys.} {\bf 29} 131.

\bibitem{Ket} Ketsaris A A 1999 {\it Turns and Special Relativity
Transformations} (submitted to {\it J. Phys. A: Math. Gen.}).

\endbib
\end{document}